\documentclass[10pt,times,a4paper,floatfix]{iopart}
\usepackage{epsfig}
\usepackage{graphics}
\usepackage{graphicx}
\usepackage{bm}
\usepackage{iopams}
\usepackage[dvipsnames]{xcolor}
\usepackage{bm}
\usepackage{times}
\usepackage{xspace}
\usepackage[varg]{txfonts}
\usepackage[normalem]{ulem} % To get strikethrough (\sout)
\usepackage[colorlinks]{hyperref}
\usepackage[caption=false]{subfig}
\usepackage{booktabs}

\definecolor{LinkColor}{rgb}{0.75, 0, 0}
\definecolor{CiteColor}{rgb}{0, 0.5, 0.5}
\definecolor{UrlColor}{rgb}{0, 0, 0.75}
\hypersetup{linkcolor=LinkColor}
\hypersetup{citecolor=CiteColor}
\hypersetup{urlcolor=UrlColor}
\usepackage{perpage}
\MakePerPage{footnote}

\begin{document}
\title{GW150914 peak frequency: a novel consistency test of strong-field General Relativity}
\author{Gregorio   Carullo$^{1}$, Gunnar 	   Riemenschneider$^{2,5}$,
Ka~Wa~     Tsang$^{3}$, Alessandro Nagar$^{4,5,6}$, Walter Del Pozzo$^{1,7}$}

\address{${}^1$ Dipartimento di Fisica ``Enrico Fermi'', Universit\`a di Pisa, Pisa I-56127, Italy}
\address{${}^2$ Dipartimento di Fisica, Universit\`a di Torino, via P. Giuria 1, I-10125 Torino, Italy}
\address{${}^3$ Nikhef -- National Institute for Subatomic Physics, Science Park, 1098 XG Amsterdam,The Netherlands}
\address{${}^4$ Centro Fermi - Museo Storico della Fisica e Centro Studi e Ricerche Enrico Fermi, Rome, Italy}
\address{${}^5$ INFN Sezione di Torino, Via P. Giuria 1, 10125 Torino, Italy}
\address{${}^6$ Institut des Hautes Etudes Scientifiques, 91440 Bures-sur-Yvette, France}
\address{${}^7$ INFN sezione di Pisa, Pisa I-56127, Italy}

\ead{gregorio.carullo@ligo.org}

\date{\today}

\begin{abstract}
We introduce a novel test of General Relativity in the strong-field regime of a binary black hole coalescence. Combining 
information coming from Numerical Relativity simulations of coalescing black hole binaries with a Bayesian 
reconstruction of the gravitational wave signal detected in LIGO-Virgo interferometric data, allows one to test theoretical 
predictions for the instantaneous gravitational wave frequency measured at the peak of the gravitational wave signal amplitude. 
We present the construction of such a test and apply it on the first gravitational wave event detected by the LIGO and Virgo Collaborations,
GW150914. The $p$-value obtained is $p=0.48$, to be contrasted with an expected value of $p=0.5$,
so that no signs of violations from General Relativity were detected.
\end{abstract}

\maketitle

%-------------------------------------------
% Introduction.
%-------------------------------------------

\textit{Introduction --} 
The detection of gravitational waves (GWs) by the LIGO and Virgo Collaborations
\cite{TheVirgo:2014hva, Abbott2018} has opened several new routes to explore observationally 
the genuine strong-field dynamics of gravity. GW150914~\cite{GW150914}
and subsequent detections \cite{GW151226,TheLIGOScientific:2016pea,Abbott:2017vtc,Abbott:2017gyy,GW170814,GW170817}
have already provided the best dynamical constraints on general relativity (GR)
\cite{TGR_paper_GW150914,TheLIGOScientific:2016pea,Abbott:2017vtc,GW170814}.
Thanks to its mass and loudness, GW150914 allowed exquisite tests of GR~\cite{TGR_paper_GW150914}
as well as the measurements of the frequency and damping time
of the least damped quasi-normal mode (QNM) of the presumed
remnant black hole (BH) resulting from a binary black hole (BBH) merger~\cite{TGR_paper_GW150914}.

GR predicts that GWs of astrophysical interest have two independent polarizations,
namely $(h_+, h_{\times})$. In GR, the instantaneous frequency of the gravitational radiation
emitted by a binary is an observable which, given a model for the emission,
can be computed as
\begin{center}
\begin{equation}\label{eq:f_t}
f(t) = \frac{1}{2\pi} \frac{d}{dt}(\arg(h_+ - i h_{\times})).
\end{equation}
\end{center}
Of particular interest is $f^{\rm peak}$, the frequency of the signal at the time at which the strain amplitude,
defined here as $\left(h_+^{2}+h_{\times}^{2}\right)^{1/2}$, reaches its peak.
This stage approximately coincides with the time when the two black holes merge,
so that $f^{\rm peak}$ encodes information on the behavior of the system in the genuine
strong-field regime. The access to the precise value of $f^{\rm peak}$ for BBHs
relies on full numerical relativity (NR) simulations. NR simulations are, however,
numerically expensive, so that they cannot be used to densely sample the 7-dimensional
binary parameter space\footnote{The mass ratio and the two three-dimensional spin components.}.
Especially in the case of spin-aligned (nonprecessing) BBHs, where more NR simulations
are available, it is however possible to suitably fit the NR data so to
obtain closed-form analytical representation of $f^{\rm peak}$ as a function
of the physical parameters (mass ratio and spins) of the coalescing
binary~\cite{Nagar:2018zoe,Bohe_f22,Healy_2014_f22,Healy:2017mvh,Healy_2018_f22,Cotesta:2018fcv}
that are viable all over the parameter space.
In this \textit{Letter}, we compare the aforementioned GR prediction
with the agnostic measurement coming from an unmodelled reconstruction
of the $(h_+ ,h_{\times})$ polarizations\footnote{More generic polarization contents can also be tested, 
see ~\cite{Isi:2017fbj}}, providing a new statistical test of GR in its strong-field regime,
similar to what was suggested in \cite{Healy:2017mvh}.

This \textit{Letter} is organized as follows: (i) we discuss our model for $f^{\rm peak}$
constructed from  full numerical solutions of Einstein's equations; 
(ii) we reconstruct, using BayesWave~\cite{Cornish:2014kda,Millhouse:2018dgi}, the value of $f^{\rm peak}$
for GW150914 in a model independent way; 
(iii) we compare the agnostic reconstruction of $f^{\rm peak}$ with the prediction
coming from a binary black hole merger in GR. Our analysis finds no evidence
for a violation of GR predictions in GW150914. 
Finally, we conclude with a discussion of future prospects.

%-------------------------------------------
% Theoretical section: fits.
%-------------------------------------------
\paragraph{The merger frequency in GR --}

Numerical relativity~\cite{GT_catalog,SXS_catalog2,Bruegmann:2006at,Husa:2007hp}
is able to simulate coalescing BBHs from several inspiral orbits through plunge,
merger and ringdown. The physical waveform is given, in the source frame, as
a multipolar expansion of the form
\begin{equation}\label{eq:multipole}
h_+ - {\rm i} h_{\times} = \sum_{\ell,m} h_{\ell m}(\theta) {}_{-2}Y_{\ell m}(\iota, \phi) \ .
\end{equation}
Here, $\theta$ collectively indicate the intrinsic physical parameters  that 
characterize the source (e.g. component masses and spin vectors), 
${}_{-2}Y_{\ell m}(\iota, \phi)$ are the $s=-2$ spin-weighted spherical
harmonics, where $(\iota, \phi)$ indicate the polar (with respect to the direction
of the orbital angular momentum) and azimuthal angle in the source frame.
The complex waveform multipoles $h_{\ell m}$ are decomposed in amplitude
and phase as 
\begin{equation}
h_{\ell m}(\theta) \equiv A_{\ell m}(\theta) e^{-i \varphi_{\ell m}(\theta)}\,.
\end{equation}
In what follows, we shall restrict our attention only to BBH systems where
the individual spins are aligned (or anti-aligned) with the orbital angular 
momentum, therefore neglecting precession effects. The description of 
the merger and postmerger regime in semi-analytic waveform models 
relies on fits of quantities extracted from NR waveform data at (or around) 
the peak of the $h_{22}$ mode 
(see e.g.~\cite{Bohe_f22,Healy_2014_f22,Healy_2018_f22,Nagar:2018zoe}).
In particular, it is possible to construct analytical representations of the frequency
of  the $h_{22} (\theta)$ mode at the peak  of $|h_{22} (\theta)|$, that we will
denote  as $f_{22}^{\rm peak}$. It is important to note that $f_{22}^{\rm peak}$
is in general different from $f^{\rm peak}$ as defined in \ref{eq:f_t}, 
which is instead computed at the peak of $\sqrt{h_+^{2}+h_{\times}^{2}}\,$.
Since the quantity that can be extracted from the interferometric data is $f^{\rm peak}$,
we will make use of an approximation, (see below) in which the two quantities are directly related and
the prediction from numerical simulations can thus be directly exploited to test the data for GR violations.

Our aim being to perform a statistical test on GW150914, it will suffice to consider 
a gravitational wave model which, in a spherical decomposition, includes only the $(\ell,m) = \{(2, 2), (2,-2)\}$ modes~\cite{Abbott:2016izl},
and that reads
\begin{equation}\label{eq:multipole2}
h_+ - i h_{\times} = {}_{-2}{Y_{22}}(\iota, \phi) h_{22}  + {}_{-2}{Y_{2-2}}(\iota, \phi) h_{2-2} .
\end{equation}
Being the posterior distribution of the inclination angle for GW150914 strongly peaked towards 
the face-off value (angular momentum of the binary pointing in the opposite direction with respect 
to the line of sight)~\cite{PE_paper_GW150914},  we can simplify the computation further by 
making the approximation that the binary was face-off with respect to the earth
(see below for a discussion of the error introduced by this approximation).
In the face-off case, one has ${}_{-2}{Y_{22}}(\pi, \phi) = 0$, and
${}_{-2}{Y_{2-2}}(\pi, \phi) = \sqrt{\frac{5}{4\pi}} e^{-2i\phi} \equiv \kappa \, e^{-2i\phi}$,
which yields
\begin{equation}\label{eq:foa}
h_+ - i h_{\times} = A_{22}(t) e^{i\varphi_{22}(t)} \, \kappa \, e^{-2i\phi},
\end{equation}
so to finally obtain
\begin{equation}
\label{eq:frequency}
f(t) = \frac{1}{2\pi} \frac{d(\arg(h_+ - i h_{\times}))}{dt} = \frac{\dot{\varphi}_{22}(t)}{2\pi} = f_{22}(t)\,,
\end{equation}
where we applied the fact that for spin-aligned binaries one has $h_{\ell,-m} = (-1)^{\ell}h^*_{\ell m}$.
Eq.(~\ref{eq:frequency}) allows to directly compare a prediction for $f^{\rm peak}_{22}$ 
to the measured peak frequency from interferometric data, $f^{\rm peak}$.
Note that, in the face-off approximation, the amplitude is given simply by:
\begin{equation}\label{eq:ampl}
| h_+ - i h_{\times} |  = A_{22}(t) \, \kappa\,,
\end{equation}
\textit{i.e.} the amplitude of the full waveform is proportional to the amplitude of the $(\ell,m)=(2,2)$ mode, 
with no mode mixing. Eq.~(\ref{eq:ampl}) implies that the peak of the amplitude of the $(\ell,m) = (2,2)$
mode will coincide with the peak of the full waveform ~\ref{eq:multipole2}.
These simplifications do not hold outside of the face-off/on approximation or whenever
subdominant multipoles are not negligible.

Using a subset of NR simulations \cite{SXS_catalog, Blackman:2015pia} with mass and spin parameters compatible with 
the released samples of GW150914 by the LVC collaboration~\cite{PE_paper_GW150914, TheLIGOScientific:2016pea, O1_samples_release}, 
we assessed the error introduced by the face-off approximation for GW150914-like signals. 
The waveforms extracted from the simulations were evaluated at the detectors location using GW150914 samples 
for the orbital angles. The peak of the frequency for each samples was extracted
first using only the ($\ell=2, m=-2$) mode and then using all the modes up to $\ell=4$.
We find the difference due to the two different estimations to be smaller than the statistical error due to the 
unmodelled waveform reconstruction (see next section), proving that, 
for our current purposes, the face-off approximation for GW150914 is valid.

Let us now turn to discuss our NR-informed analytical model
for the merger frequency. The NR frequency data (see below)
of the $h_{22}$ multipole are fitted with the following functional ansatz
\begin{equation}
\label{eq:fttfit}
f_{22}^{\rm peak} 	= \frac{1}{2\pi M} f_{22}^{\nu=0} f_{22}^{\rm orb}(\nu)f_{22}^{\hat{S}}(\hat{S},X_{12})\ ,
\end{equation}
where $\nu\equiv m_1m_2/M^2$ is the symmetric mass ratio, $M\equiv m_1+m_2$ the total mass of the binary, 
$X_{12}\equiv (m_1-m_2)/M=\sqrt{1-4\nu}$, with $m_1\geq m_2$ and $\hat{S}\equiv (S_1+S_2)/M^2$ 
is the mass-normalized total spin of the system.  Ref.~\cite{Nagar:2018zoe} illustrated
that the use of $\hat{S}$ allows for a rather natural, and simple, representation of the peak
frequency all over the currently NR-covered parameter space of spin-aligned BBHs coalescences.
In Eq.~(\ref{eq:fttfit}), $f^{\nu=0}$ refers to the value of $f^{\rm peak}_{22}$ from the merger in
the extreme-mass-ratio limit, obtained by numerically solving the Teukolsky equation with
a point-mass source~Ref.~\cite{Harms:2014dqa,Nagar:2018zoe}. 
The nonspinning factor, $f_{22}^{\rm orb}(\nu)$, is represented as
\begin{equation}
f_{22}^{\rm orb}(\nu) = 1 + a_1\nu + a_2\nu^2\,.
\end{equation}
The value of the coefficients $a_1$ and $a_2$ is determined from 19 non-spinning, SXS waveforms 
with mass ratios $1\leq m_1/m_2\leq 10$~\cite{SXS_catalog2}. The (equal mass) spin-dependence 
is modeled as a rational function of the form
\begin{equation}
\label{eq:f0}
f_{22}^{\hat{S}}(\hat{S},X_{12}=0)=\frac{1+b^{m_1=m_2}_1\hat{S}+b^{m_1=m_2}_2\hat{S}^2}{1+b^{m_1=m_2}_3\hat{S}},
\end{equation}
and informed by 39 equal-mass, spin-aligned, waveforms from the SXS 
collaboration~\cite{SXS:catalog,Chu:2009md,Lovelace:2011nu,Buchman:2012dw,Mroue:2013xna,Hemberger:2013hsa,Scheel:2014ina,Blackman:2015pia,Chu:2015kft,Lovelace:2016uwp}.
To incorporate the additional dependence on (symmetric) mass ratio, we introduce
it through $X_{12}$ by replacing the equal-mass coefficients in Eq.(~\ref{eq:f0})
with 
\begin{equation}
\label{eq:expol1}
b^{m_1=m_2}_1 \rightarrow  \frac{b^{m_1=m_2}_1 + c_1X_{12}}{1+c_2X_{12}} \ ,
\end{equation}
\begin{equation}
\label{eq:expol2}
b^{m_1=m_2}_3 \rightarrow  \frac{b^{m_1=m_2}_3 + c_3X_{12}}{1+c_4X_{12}} \ ,
\end{equation}
where the additional coefficients $c_i$ are then fitted using test-particle data, 77 additional 
SXS spinning waveforms and 14~additional NR waveforms obtained using the 
BAM~code\cite{Bruegmann:2006at,Gonzalez:2006md,Husa:2015iqa} 
waveforms. The coefficients are given explicitly in Table~\ref{tab:freq_coefficients}. 
This model is an improvement of  the corresponding one presented in Appendix F 
of Ref.~\cite{Nagar:2018zoe} in three aspects: (i) the non-spinning, test-particle behavior  
is imposed explicitly; (ii) a quadratic dependence on $\hat{S}$, informed by equal-mass data, 
is introduced in the numerator of $f_{22}^{\hat{S}}$; (iii) rational functions, instead of 
quadratic polynomials, are used in the extrapolation of the spinning behavior from 
the equal mass case, see Eq.(~\ref{eq:expol1},\ref{eq:expol2}). 

%===============================
% Table of the fit coefficients
%===============================
\begin{table}
\centering
  \caption{\label{tab:freq_coefficients} Explicit coefficients and their errors for the merger frequency fit.
  All coefficients and errors where obtained with the function \texttt{fitnlm} of \texttt{matlab}.
  The analytic template of the fit is defined in Eq.(~\ref{eq:fttfit},\ref{eq:expol2}).}

\begin{tabular}{c c c l l c}
\hline\hline
 &$f_{22}^{\nu=0}$ & $=$ & $\;\;\,0.273356$ & &\\
 &$a_1$ & $=$ & $\;\;\,0.84074$ & $\pm 0.014341$  &\\
 &$a_2$ & $=$ & $\;\;\,1.6976$ & $\pm 0.075488$  &\\ \hline
 &$b^{m_1=m_2}_1$ & $=$ & $-0.42311$ & $\pm 0.088583$  &\\
 &$b^{m_1=m_2}_2$ & $=$ & $-0.066699$ & $\pm 0.042978$  &\\ 
 &$b^{m_1=m_2}_3$ & $=$ & $-0.83053$ & $\pm 0.084516$  &\\ \hline
 &$c_1$ & $=$ & $\;\;\,0.15873$ & $\pm 0.1103$  &\\
 &$c_2$ & $=$ & $-0.43361$ & $\pm 0.2393$  &\\ 
 &$c_3$ & $=$ & $\;\;\,0.60589$ & $\pm 0.076215$  &\\ 
 &$c_4$ & $=$ & $-0.71383$ & $\pm 0.096828$  &\\ \hline\hline
\end{tabular}
\end{table}

%=============================================================
% Quick discussion of what to do for the generic case.
%=============================================================

Although for the GW150914 case the face-off approximation is valid, thus simplifying the problem considerably, 
generic BBH event will require an extension of the present work, including contributions of subdominant modes.
We leave such an extension (and the exploration of its observational consequences in the upcoming runs of interferometric 
observatories) to future work. Nevertheless, let us mention two routes to faithfully represent the dependence on $(\iota,\phi)$, 
the polar and azimuthal angles, for systems where the face-off approximation is invalid: 
(i)  The peak frequency could be reconstructed directly using an 
NR-surrogate model~\cite{Varma:2018mmi,Blackman:2017dfb} or a waveform 
model containing higher mode contributions~\cite{London:2017bcn,Cotesta:2018fcv,Mehta:2019wxm,Husa_IMRPhenom}.  
Or (ii) the peak frequency could be fitted directly to NR data, extending the procedure discussed above to subdominant modes~\cite{Bruegmann:2006at,Gonzalez:2006md,Husa:2015iqa,SXS:catalog,Chu:2009md,Lovelace:2011nu,Buchman:2012dw,Mroue:2013xna,Hemberger:2013hsa,Scheel:2014ina,Blackman:2015pia,Chu:2015kft,Lovelace:2016uwp,Harms:2014dqa}.

\begin{figure}[t]
\center
\includegraphics[width=0.5\textwidth]{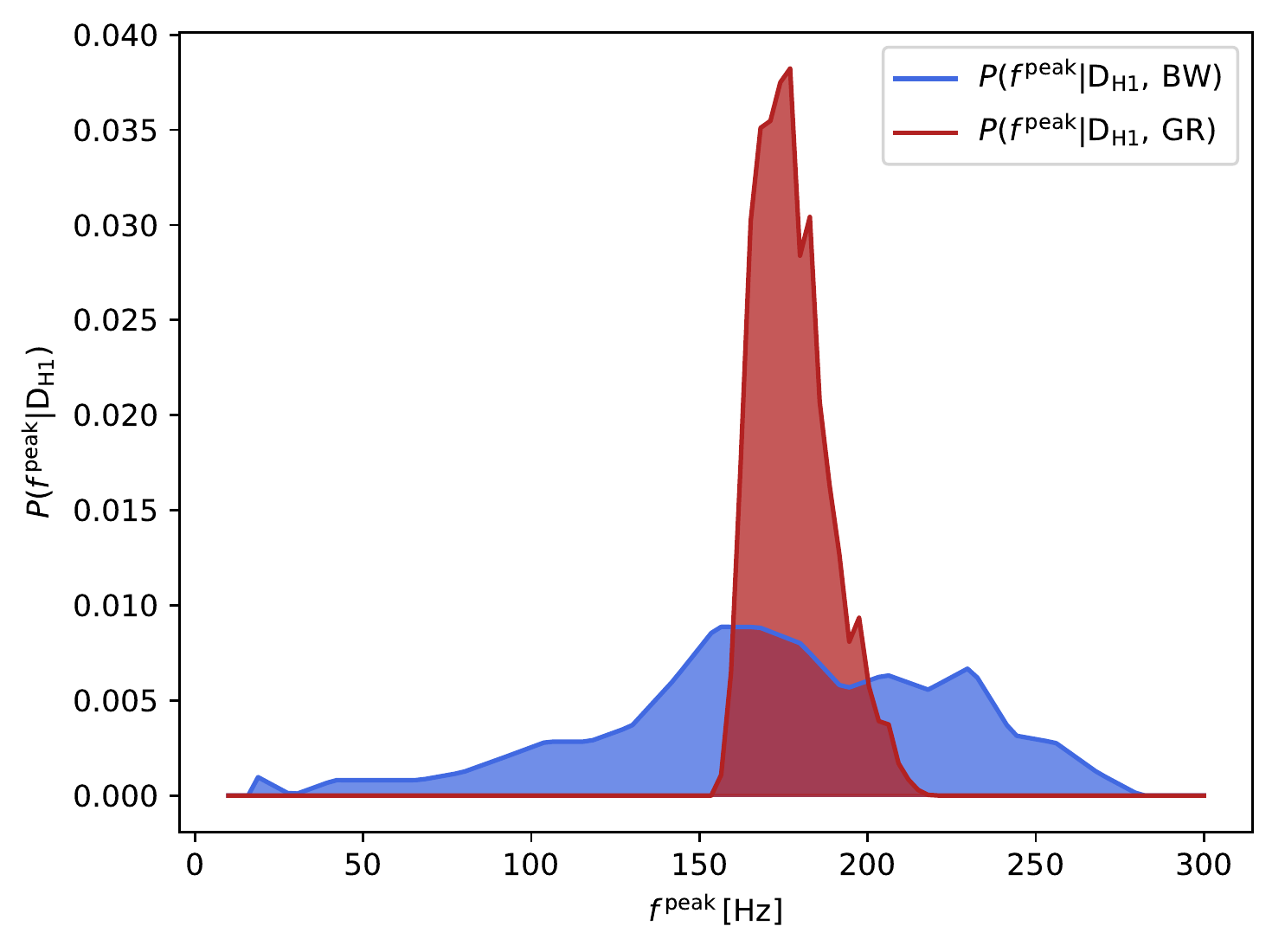}
\caption{\label{fig:f22_GW} Peak frequency obtained by evaluating the GR prediction, 
Eq.(~\ref{eq:fttfit}), on the samples of GW150914 (red) and by the unmodelled \texttt{BayesWave} reconstruction (blue) using Hanford data.}
\end{figure}

\textit{Reconstructed peak frequency --} 
Having at our disposal a GR prediction for $f_{22}^{\rm peak}$, 
we now proceed to evaluate the fit presented in Eq.~(\ref{eq:fttfit}) 
on the measured physical parameters of the system. Rather than using point estimates, 
we estimate the prediction for the frequency $f^{\rm peak}$
by evaluating it over the posterior distribution of the physical parameters of the system (masses and spins) obtained 
from a full bayesian parameter estimation analysis.
We use the publicly available posterior samples from the LIGO/Virgo collaboration~\cite{PE_paper_GW150914, TheLIGOScientific:2016pea, O1_samples_release}.
The template used to perform this analysis was IMRPhenomPv2~\cite{Husa_IMRPhenom}.
From the application of Eq.~(\ref{eq:fttfit}) on the GW150914 posterior samples, we obtain a distribution for  
$f^{\rm peak}$ as predicted by GR, shown in Figure~\ref{fig:f22_GW}.

%-----------------------
% Representative plot.
%-----------------------

%----------

%-------------------------------------------
% GR agnostic evaluation of frequency: BayesWave computation.
%-------------------------------------------

In order to obtain an evaluation of the peak frequency 
which is not tied to the GR predicted phase evolution, 
we run the \texttt{BayesWave} algorithm on GW150914, obtaining 
a distribution of reconstructed waveforms for the given dataset.
\texttt{BayesWave} (\cite{Cornish:2014kda,Millhouse:2018dgi}) is a morphology-independent search
algorithm which can distinguish signal hypothesis from glitch or noise hypotheses
by comparing the corresponding Bayesian evidences and give the associated waveform reconstruction.
The $h_+$ polarization is decomposed into a sum of Morlet-Gabor wavelets, which are a complete set and hence allowing 
the unmodelled reconstruction of any possible gravitational-wave signal without making any assumption on its morphology.
The number and all intrinsic parameters of wavelets (amplitude, central frequency, decay time, central time, phase offset) 
are sampled by means of a trans-dimensional Reversible Jump Markov Chain Monte Carlo algorithm.
$h_\times$ is then built as $h_\times = \epsilon h_+ e^{i\pi/2}$ with $\epsilon$ an ellipticity parameter.
Using the reconstructed waveform, 
we obtain a distribution of the instantaneous frequency,
evaluated at the merger time using Eq.(~\eref{eq:f_t}).
To take into account the uncertainties on $t_{\rm peak}$ due to the finite 
sampling rate and the face-off approximation, we additionally marginalize the posterior 
distributions over $t_{\rm peak}$ uncertainty.
The result of such a computation is shown in Figure~\ref{fig:f22_GW}.
To test our method, we performed both our modelled and unmodelled frequency reconstruction on a set of simulated data added to gaussian noise.
The orbital parameters and source localisation were similar to the ones of GW150914, so that all our assumptions were still valid.
We obtained results completely consistent with the ones reported below.

Combining the peak frequency distribution obtained by exploiting 
the GR prediction and the model-agnostic distribution
of the frequency reconstructed through {\tt BayesWave},
we can now build the null variable
\begin{equation}
\Delta f \equiv f^{\rm peak}_{\rm rec}-f^{\rm peak}_{\rm GR}. 
\end{equation}
Define the posteriors $p(f^{\rm peak}|\mathrm{D}\, ,\mathrm{GR})\equiv q(f^{\rm peak})$ and 
$p(f^{\rm peak}|\mathrm{D}\, ,\mathrm{BW})\equiv r(f^{peak})$.
Since $\Delta f$ is the difference of two independently distributed random variables, 
its posterior distribution can be computed as
\begin{equation}
g(\Delta f) = \int{r(f^{\rm peak}_{rec}) \, q(f^{\rm peak}_{rec}-\Delta f) df^{\rm peak}_{rec}}\, .
\end{equation}
Under the assumption that GR is the correct theory describing gravitational 
interactions (the \textit{null} hypothesis), $g(\Delta f)$ must to be centered 
around 0. As a quantitative indicator of the agreement between GR and the observed value of $\Delta f$, 
we compute the cumulative distribution of the null variable $G(\Delta f) = \int_{-\infty}^{\Delta f} g(x)dx$ 
and use it to calculate a p-value: $p\equiv\mathrm{min\left[1-G(0),G(0)\right]}$. 

\begin{figure}[t]
\center
\includegraphics[width=0.5\textwidth]{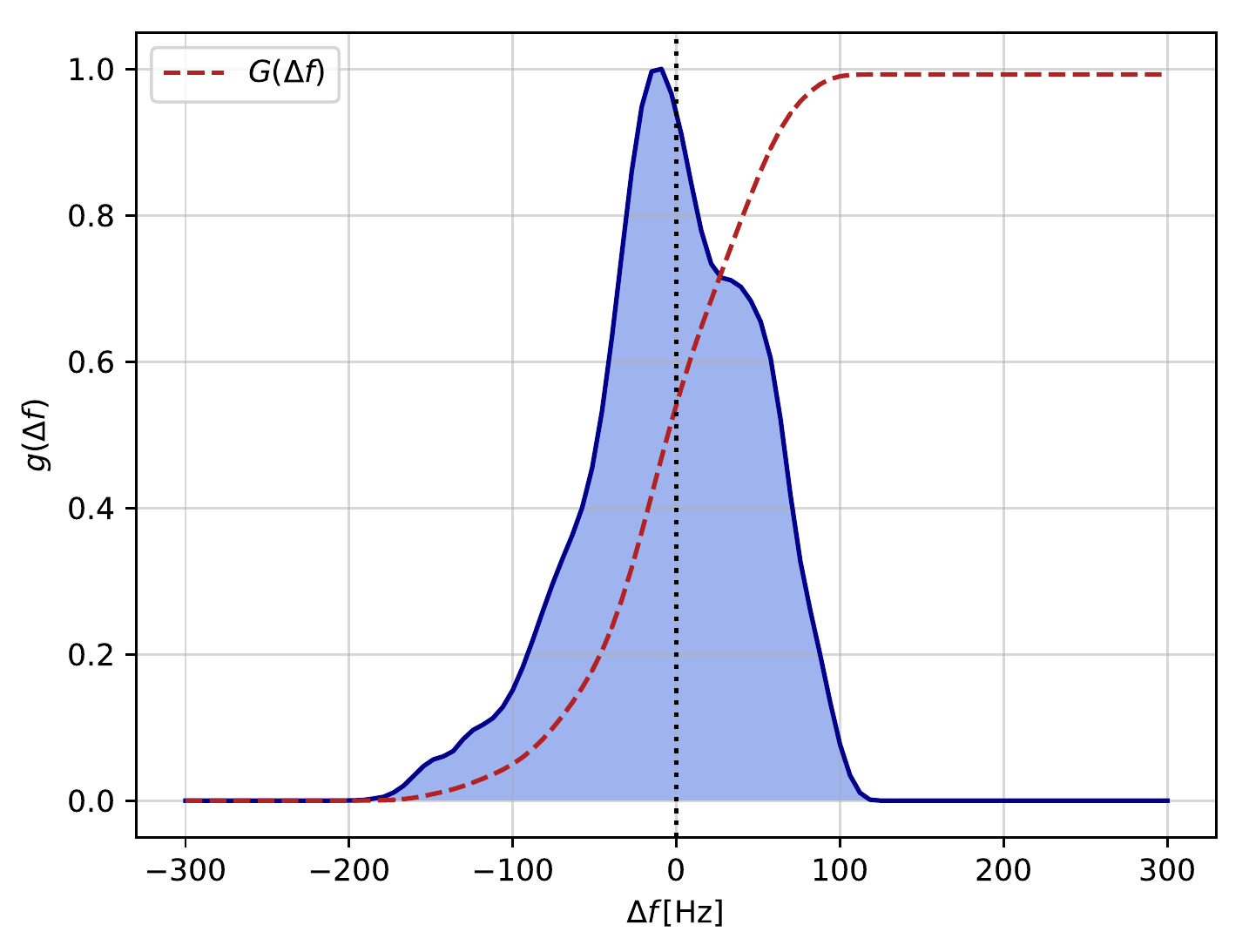}
\caption{\label{fig:diff_H1} Distribution of the difference between the theoretical prediction and 
unmodeled evaluation of the peak frequency using Hanford data (blue). The cumulative distribution is shown by the red dashed line.}
\end{figure}

The result of this calculation using the Hanford strain, Fig.~\ref{fig:diff_H1}, 
shows no significant deviation from GR predictions, yielding a p-value: $p = \mathrm{0.48}$. 
The equivalent calculation using the Livingstone strain yields: $p = 0.46$.
Under the null hypothesis we would expect the p-value to be $p = 0.5$, 
thus the data show no evidence for GR violations.

%-----------------------
% Representative plot.
%-----------------------

%----------

%-------------------------------------------
% Conclusions.
%-------------------------------------------

\textit{Discussion and conclusions --} In this \textit{Letter}, we exploited the comparison between the 
theoretical predictions for the value of the instantaneous frequency of gravitational waves 
at the peak of the waveform emitted by a BBH coalescence, with its data-driven reconstruction 
to perform a new genuinely strong-field test of GR. This frequency in fact approximately corresponds
to the instant when the two black holes merge. We investigated the case for GW150914 and 
found no evidence for violations of GR. 

Our work is the first test of this kind and it is targeted at GW150914. Therefore, 
we relied on a set of approximations that are valid only for GW150914. 
Relaxing the face-off/on approximation, for instance, implies that the time shift between 
the peak of the $(\ell,m) = (2,2)$ mode and the full multipolar waveform has to be taken into account,
together with the relative phase of the different modes, which will be dependent on the 
orientation angles in the source frame ($\iota, \phi$).
Thus, in the general case, a comparison of the equations used in this paper 
with the data will not be directly possible. To do so, one will need to include 
higher modes (e.g. similarly fitted to NR data as for the $\ell=m=2$ one) 
and the relative phases between the modes.
The other main limitation of our test is certainly the accuracy with which $f^{\rm peak}_{\rm rec}$ can be currently reconstructed.
The statistical sensitivity of the test will greatly improve once the LIGO-Virgo network will come back online
at enhanced sensitivity and more events, with clearly identifiable waveform peaks, will be detected.
The sensitivity of the test can be improved by combining results
from different detectors for a given event, together with combining results from different events 
under the null hypothesis. We leave such extensions of the current test to future work.

%\begin{acknowledgments}
\textit{Acknowledgments}
The authors would like to thank the GeorgiaTech Numerical Relativity division for sparkling interest into this problem
and Tyson Littenberg for his assistance in using BayesWave.
K.W.T. ~is supported by the research program of the Netherlands Organisation for Scientific Research (NWO).
W.D.P. is funded by the ``Rientro dei Cervelli Rita Levi Montalcini'' Grant of the Italian MIUR.
This work greatly benefited from discussions within the \textit{strong-field} working group
of the LIGO/Virgo collaboration. This research has made use of data, software and/or
web tools obtained from the Gravitational Wave Open Science Center (https://www.gw-openscience.org), 
a service of LIGO Laboratory, the LIGO Scientific Collaboration and the Virgo Collaboration. 
LIGO is funded by the U.S. National Science Foundation. Virgo is funded by the 
Italian Istituto Nazionale di Fisica Nucleare (INFN), the French Centre National de Recherche Scientifique (CNRS) 
and the Dutch Nikhef, with contributions by Polish and Hungarian institutes.
%\end{acknowledgments}

\textit{References}
\bibliographystyle{iopart-num}
\bibliography{Merger_freq_consistency_test_Bibliography}
%% +++++++++++++++++++++++++++++++++++++++++++++++++++++++
\end{document}